\documentclass[12pt]{iopart}

\usepackage{graphicx}
\usepackage{color}
\usepackage{epsfig}
\usepackage{float}
\usepackage{multirow}
\expandafter\let\csname equation*\endcsname\relax
\expandafter\let\csname endequation*\endcsname\relax
\usepackage{amsfonts}
\usepackage{amssymb}
\usepackage{amsmath}
\usepackage{tabularx,url,color}
\usepackage{hyperref}

\usepackage{lineno}
\usepackage{upgreek}
\usepackage{comment}

\newcommand*\rot{\rotatebox{90}}

\newcommand{\mus}  {$\upmu \text{s}$}

\input{definitions}

\begin{document}
\leftline{Dated: \today}

\title{Calibration of Advanced Virgo and Reconstruction of the Gravitational Wave Signal h(t) during the Observing Run O2}

\author{F.~Acernese$^{1,2}$, 
T.~Adams$^{3}$, 
K.~Agatsuma$^{4}$, 
L.~Aiello$^{5,6}$, 
A.~Allocca$^{7,8}$, 
M.~A.~Aloy$^{9}$, 
A.~Amato$^{10}$, 
S.~Antier$^{11}$, 
M.~Ar\`ene$^{12}$, 
N.~Arnaud$^{11,13}$
% ...
\\
(Virgo~Collaboration)\\
S.~Kandhasamy$^{L1}$,  
A.~L.~Urban$^{L2}$     
}

{$^{\dag}$Deceased, February 2018; {$^{\ddag}$Deceased, November 2017.}\\

\address{$^{1}$Universit\`a di Salerno, Fisciano, I-84084 Salerno, Italy}
\address{$^{2}$INFN, Sezione di Napoli, Complesso Universitario di Monte S.Angelo, I-80126 Napoli, Italy}
\address{$^{3}$Laboratoire d'Annecy de Physique des Particules (LAPP), Univ. Grenoble Alpes, Universit\'e Savoie Mont Blanc, CNRS/IN2P3, F-74941 Annecy, France}
\address{$^{4}$Nikhef, Science Park 105, 1098 XG Amsterdam, The Netherlands}
\address{$^{5}$Gran Sasso Science Institute (GSSI), I-67100 L'Aquila, Italy}
\address{$^{6}$INFN, Laboratori Nazionali del Gran Sasso, I-67100 Assergi, Italy}
\address{$^{7}$Universit\`a di Pisa, I-56127 Pisa, Italy}
\address{$^{8}$INFN, Sezione di Pisa, I-56127 Pisa, Italy}
\address{$^{9}$Departamento de Astronom\'{\i}a y Astrof\'{\i}sica, Universitat de Val\`encia, E-46100 Burjassot, Val\`encia, Spain}
\address{$^{10}$Laboratoire des Mat\'eriaux Avanc\'es (LMA), CNRS/IN2P3, F-69622 Villeurbanne, France}
\address{$^{11}$LAL, Univ. Paris-Sud, CNRS/IN2P3, Universit\'e Paris-Saclay, F-91898 Orsay, France}
\address{$^{12}$APC, AstroParticule et Cosmologie, Universit\'e Paris Diderot, CNRS/IN2P3, CEA/Irfu, Observatoire de Paris, Sorbonne Paris Cit\'e, F-75205 Paris Cedex 13, France}
\address{$^{13}$European Gravitational Observatory (EGO), I-56021 Cascina, Pisa, Italy}
% ...
\address{$^{L1}$LIGO Livingston Observatory, Livingston, LA 70754, USA}            
\address{$^{L2}$LIGO, California Institute of Technology, Pasadena, CA 91125, USA} 
\date{\today}

\begin{abstract}
In August 2017, Advanced Virgo joined Advanced LIGO for the end of the~O2 run, leading to the first gravitational waves
detections with the three-detector network.
This paper describes the Advanced Virgo calibration and the gravitational wave strain $h(t)$ reconstruction during~O2.
The methods are the same as the ones developed for the initial Virgo detector and have already been described in previous publications;
this paper summarizes the differences and emphasis is put on estimating systematic uncertainties.
Three versions of the $h(t)$ signal have been computed for the Virgo O2 run, an online version and two post-run reprocessed versions
with improved detector calibration and reconstruction algorithm. 
A photon calibrator has been used to establish the sign of $h(t)$ 
and to make an independent partial cross-check of the systematic uncertainties.
The uncertainties reached for the latest $h(t)$ version are $5.1\%$ in amplitude,
$40\,\text{mrad}$ in phase and $20\,\upmu\text{s}$ in timing. 
\end{abstract}

\maketitle

\tableofcontents


\section{Introduction}

The Advanced Virgo detector~\cite{TDR,TheVirgo:2014hva} is located near Pisa (Italy) and is looking for gravitational waves  
sources emitted by astrophysical compact sources in the frequency range 10~Hz to a few~kHz.
Advanced Virgo joined the two LIGO detectors~\cite{0264-9381-32-7-074001,bib:ligoreco} at the end of the O2 observation run,
from $1^{\text{st}}$ to $26^{\text{th}}$ August 2017.
The data of all the detectors have been used together to search for gravitational wave sources.
During this run, GW170814~\cite{PhysRevLett.119.141101} was the first gravitational wave event from a binary black hole coalescence detected by the three detectors
and GW170817~\cite{PhysRevLett.119.161101,bib:GW170817_PE} was the first gravitational wave signal detected from a binary neutron star coalescence.
\ \\
The Advanced Virgo optical configuration for O2 was that of a power-recycled interferometer 
with 3~kilometer long  Fabry-Perot cavities in the arms. Signal recycling was not implemented yet.
\ \\
The gravitational wave strain couples to the longitudinal length degree of freedom of the interferometer.
But, to operate the interferometer, the relative positions of the different mirrors are controlled within tight limits~\cite{TDR}.
The control extends to a few hundred hertz and modifies the interferometer response to gravitational wave in that bandwidth.
Above a few hundred hertz, the suspended mirrors behave as free falling masses around their position:
the main effect of a passing gravitational waves is a frequency-dependent variation of the power at the output of the interferometer, 
following the interferometer optical response.
\ \\
The main purpose of the Virgo calibration is to allow to reconstruct the amplitude $h(t)$ of the gravitational wave strain
from the interferometer data.
In the long wavelength approximation, the differential length of the interferometer arms, $\Delta L = L_x - L_y$, is related to the gravitational wave strain $h$ by:
\begin{eqnarray}
h=\frac{\Delta L}{L_0}\quad\quad\text{where}\quad L_0\ = \ 3\ \text{km}
\label{eq:1}
\end{eqnarray} 
For coherent search of gravitational waves with multiple detectors, 
the sign of $h(t)$ must be well defined across detectors.
For Virgo, $L_x$ and $L_y$ respectively stand for the North and the West arm lengths.
\ \\
Generally speaking, in the $h(t)$ reconstruction, 
we remove from the dark fringe signal the contributions of the controls signals by subtracting the corrections
applied to the mirrors and we correct for the interferometer optical response transfer function.
Hence, the responses of the mirror longitudinal actuators have to be calibrated,
as well as the readout electronics of the output power and the interferometer optical response.
Absolute timing is also a critical parameter for multi-detector analysis,
in particular to determine the direction of the gravitational wave source in the sky, by using the time of flight between gravitational wave detectors.
Since the typical timing accuracy is on the order of 0.1~ms~\cite{Localization:2018_LRR}, 
absolute timing precision must be of the order of tens microseconds or less.
\ \\
The scope of this paper is to give an overview of the Advanced Virgo calibration and $h(t)$ reconstruction
during the run~O2, and to describe the systematic uncertainties.
The principle of the calibration and reconstruction is still the same as the one developed for 
the initial Virgo detector and already described in~\cite{2011CQGra..28b5005A} and~\cite{Accadia:2014fbz}.
The methods are thus not described again in detail in this paper.
In section~\ref{detector}, we briefly give an overview of the Virgo detector components that are relevant
for calibration during O2, emphasizing the differences with respect to the initial Virgo.
In section~\ref{calib}, we describe the calibration of the photodiode readout and mirror actuation.
More details are given for the procedures that were modified since initial Virgo,
and the systematic uncertainties of each calibration step are provided.
Section~\ref{recon} shows how the $h(t)$ values have been reconstructed using the parameters,
delays and transfer functions determined by the calibration. 
We also describe the different validation checks done on the reconstructed $h(t)$ channel
and how they were used to estimate the systematic uncertainties in terms of amplitude, phase and timing.
\ \\
For O2, three versions of the $h(t)$ reconstruction have been performed:
(i) the ''Online'' reconstruction used during O2 to provide $h(t)$ with a latency of about 20~s to the low-latency 
gravitational wave searches that triggered alerts to our multi-messengers partners. 
This online reconstruction was based on the calibration parameters estimated with the data taken before the start of O2.
(ii) a first reprocessing, ''V1O2Repro1A''\footnote{
V1 for Virgo detector, O2 for O2 run, ReproXY with X the reprocessing number and Y a letter to tag specific cases if needed.
} which was run in September 2017 with improved calibration parameters determined using 
calibration data taken during and after O2 and an improved reconstruction algorithm.
This version was used for the first O2 papers with Virgo data published in Fall 2017 and in~\cite{bib:GW170817_PE}.
(iii) a second reprocessing, ''V1O2Repro2A'', which was run in January 2018 with finely tuned calibration models
and further improvements of the reconstruction algorithm.
The results shown in this paper pertain to the final actuator calibration 
and to the second $h(t)$ reprocessing, except when stated otherwise.


\section{The Advanced Virgo detector during O2}\label{detector}

Most of the detector characteristics relevant to calibration and described in~\cite{2011CQGra..28b5005A,Accadia:2014fbz} for initial Virgo
were still valid for Advanced Virgo during O2.
They are briefly summarized in this section, with emphasis on the relevant modifications compared to the inital Virgo detector.

The optical configuration of Advanced Virgo, similar to initial Virgo, is a power-recycled interferometer with Fabry-Perot cavities
as shown in figure~\ref{fig:itfconfig}.
All the mirrors of the interferometer are suspended to a chain of pendulums for seismic isolation.
The input beam is provided by a Nd:YAG laser with a wavelength $\lambda = 1064\ \text{nm}$.
The power at the input of the interferometer during O2 was about 14~W.
The finesse of the 3-km long Fabry-Perot cavities in the arms has been increased to about 450.
The readout of the interferometer main output signal was changed from heterodyne detection to homodyne (or DC) detection~\cite{TheVirgo:2014hva}
but this has no impact on the calibration procedures since the used photodiode signal is proportional to 
the interferometer differential arm length in both cases.
The interferometer arm length difference is controlled to keep a destructive interference at the interferometer output port, 
with a slight offset to allow for DC detection.
In order to control the other degrees of freedom of the interferometer, the laser beam is phase modulated at a few tens~MHz 
(6, 8 and 56~MHz among others) and the error signals are acquired and demodulated from photodiodes located at various places in the interferometer.

Data from the interferometer, like the optical power of various beams and the different control signals, are times series, 
recorded at 10~kHz or 20~kHz. The data are time-stamped using the Global Positioning System (GPS).

\subsection{Mirror longitudinal actuation}
Each Virgo mirror is suspended to a complex seismic isolation system.
The bottom part is a double stage system with the so-called {\it marionette} as the first pendulum.
The mirror is suspended to the marionette by pairs of thin steel wires. 
As a modification, the reference mass is now suspended to the seismic filter 
above the marionette instead of being attached to the marionette as in initial Virgo.
This makes the mechanical response of the pendulum 
more complex. But the resonances being below 1~Hz, the mechanical response above 10~Hz still has
a simple $1/f^2$ behavior.\\

The position of the marionette and mirror is adjusted with electromagnetic actuators: 
permanent magnets are attached to the marionettes and on the back of the mirrors;
a set of coils attached to the reference mass and whose current is driven by some electronic device allows to steer the suspended objects.
The longitudinal controls are distributed between the marionette (up to a few tens of hertz) and the mirror (up to a few hundred hertz).
Therefore, the marionette and mirror actuation responses need to be measured up
to $\sim100\ \text{Hz}$ and up to $\sim1\ \text{kHz}$ respectively.\\

The actuation response includes the actuator itself and the response of the suspended mirror.
The actuator is composed of a digital part, a Digital to Analog Converter (DAC), and the analog electronics
which converts the DAC output voltage into a current flowing through the coil.


\subsection{Sensing of the interferometer output power and control loops}
The main output signal of the interferometer is the power at the dark port. It is sensed using two photodiodes.
The photodiodes and their readout electronics have been changed with respect to initial Virgo.
In particular, the main channel  $\mathcal{P}_{DC}$, that measures the output power components from 0 to 10~kHz,
is now obtained from the blend of two channels measured and digitized in two frequency bands. 
For the pick-off beam used for interferometer control, the demodulated channels, $\mathcal{P}_{AC}$, are extracted
using digital demodulation~\cite{TheVirgo:2014hva}, instead of analog demodulation in initial Virgo.
However, these modifications do not impact the calibration procedures.\\
The principle of the longitudinal control loops is the same as described in~\cite{Accadia:2014fbz}
and sketched in figure~\ref{fig:itfconfig}.


\begin{figure}[h!]
  \begin{center}
    \includegraphics[angle=0,width=0.7\linewidth]{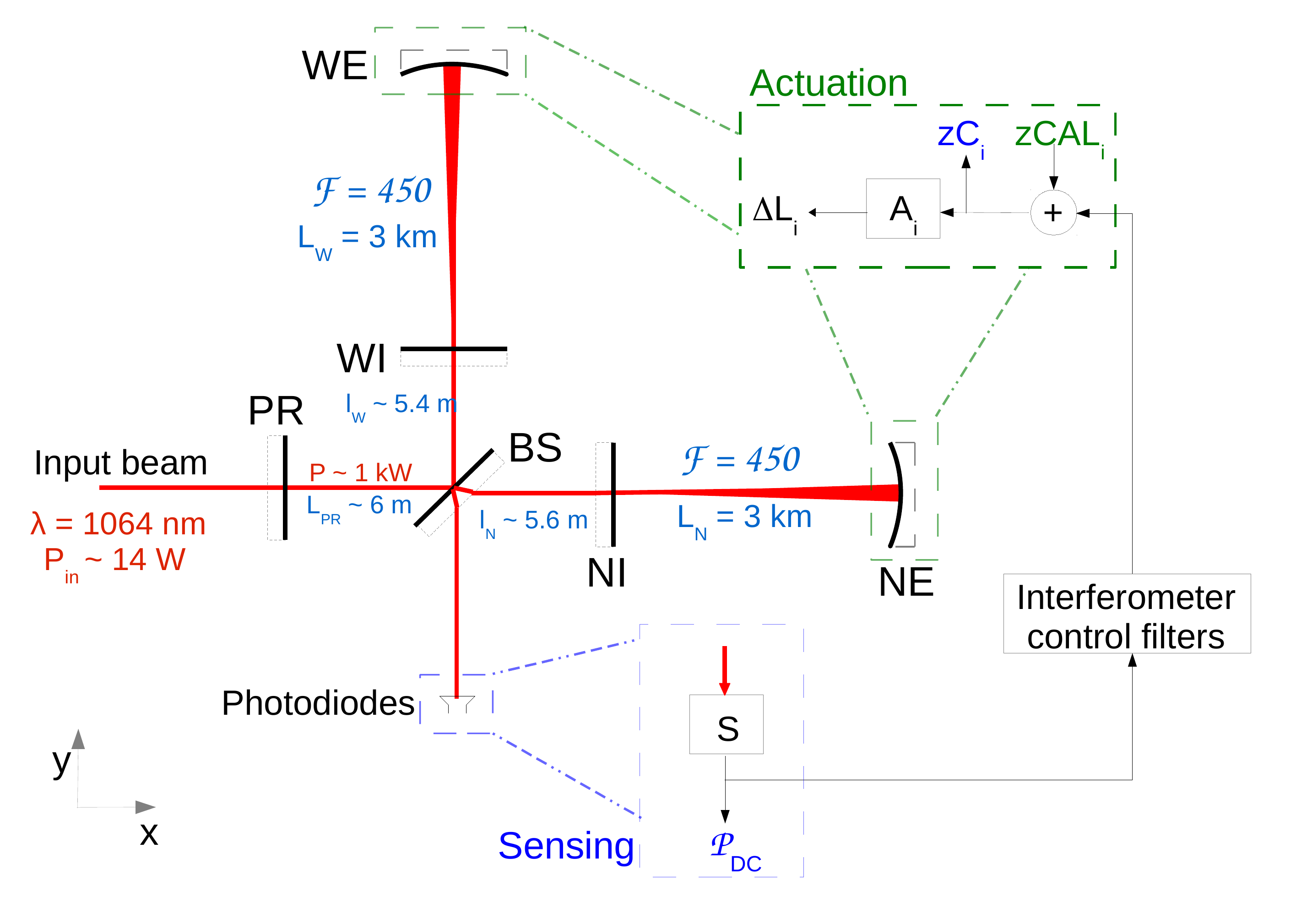}
    \caption{Configuration of the Advanced Virgo interferometer during O2 
	and sketch of the differential arm length control loop. 
    }
    \label{fig:itfconfig}
  \end{center}
\end{figure}


\section{Calibration of Advanced Virgo}
\label{calib}

To reconstruct the gravitational wave strain $h(t)$, we need to calibrate the sensing part (photodiode readout)
and the actuators (coils acting on the mirrors and on their suspensions).
More details on the procedures used can be found in the Virgo notes~\cite{calibnote_Repro1A, calibnote_Repro2A}.\\

Specific calibration data were taken a few days before Virgo joined the O2 run,
every Tuesday during the run (about two hours of calibration each week), 
and finally during a measurement campaign over several days performed just after the end of O2.
The results presented in this section are extracted from the whole dataset.

\subsection{Timing and Sensing}\label{lab:TimingAndSensing}

The photodiode readout (sensing part of the control loop) is paced by
a timing distribution system slaved to a main GPS receiver.
This receiver provides a GPS timestamp and a clock distributed to all the digital electronic
devices (ADC, DAC, real-time PCs, DSP boards, ...) to enslave their local clock
and to time the data.
\ \\
In order to estimate the absolute timing of the system, the 1~PPS (Pulse Per Second) clock signal of the current GPS receiver, 
already used for Virgo+ since 2008, has been compared to the 1~PPS of the initial Virgo GPS receiver~\cite{calibnote_timingStabilityVSR2}.
No offset between the two systems was found within uncertainties of $4$~\mus.\\
In addition, the Virgo clock has been compared to a clock provided by an independent atomic clock~\cite{calibnote_timingCheck}.
The atomic clock being free, it is expected to drift. A linear drift has been estimated over five days in September~2017, after~O2,
and is assumed to have been the same during~O2. 
After removing this linear drift, the relative variations of the Virgo clock compared to the atomic clock 
were below 13\,\mus\ during~O2. 
Note that part of such variations may still come from the atomic clock whose drift is not expected to be perfectly linear.
\ \\
The full sensing chain of the output photodiodes is described in~\cite{calibnote_timingB1}.
Using the clock signal from the main GPS receiver sent through a LED in front of the output photodiodes
and recording the signals at 1~MHz, 
we have measured the position of the 1~PPS of the clock in the Virgo data.
The measured delay of the sensing chain ($140\pm3$~\mus) was in agreement with the expected value ($142\pm1$~\mus).
\ \\
From these uncertainties on the Virgo GPS absolute timing and on the photodiode readout chain delay,
we have estimated a conservative systematic uncertainty of~20\,\mus\ on the absolute timing of the
main output signal of the interferometer, $\mathcal{P}_{DC}$.





\subsection{Calibration of the BS, NI and WI mirror actuators}

\begin{figure}[tb]
  \begin{center}
    \includegraphics[angle=0,width=0.7\linewidth]{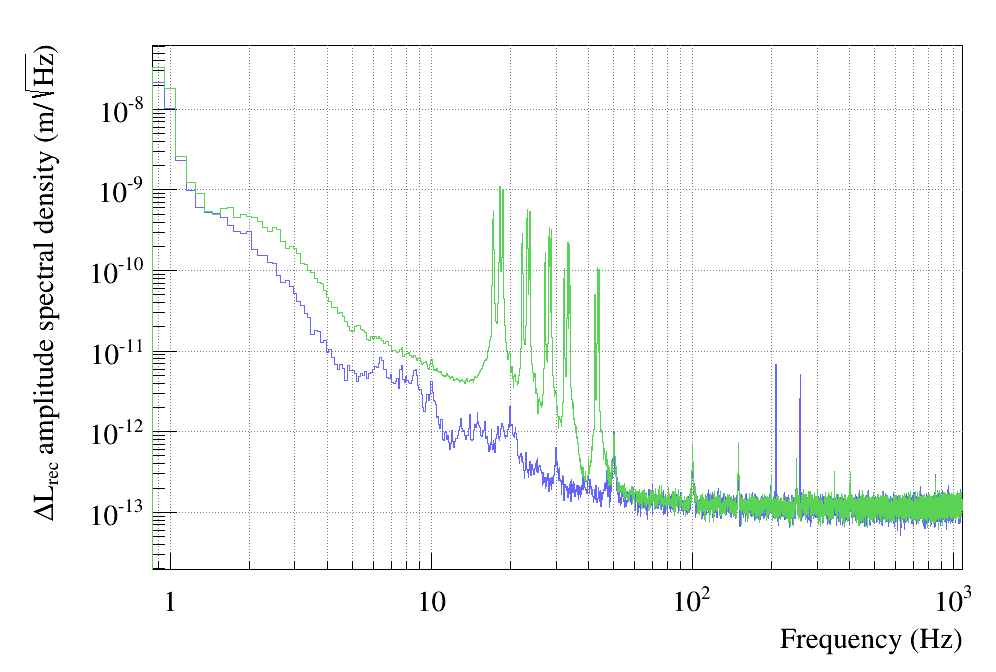} 
    \caption{Amplitude spectral density of the reconstructed $\Delta L$ signal in the free Michelson configuration.
      Two dataset are shown with different sinewaves excitations applied on the mirror actuators.
      Green: five sinewaves are applied between 15~Hz and 45~Hz to each of the three mirrors.
      Blue: two sinewaves are applied between 200~Hz and 250~Hz to each of the three mirrors.
      The plots are made from two dataset lasting for 60~s. 
    }
    \label{fig:freemich_sens}
  \end{center}
\end{figure}

The Advanced Virgo O2 calibration relies on several length references used in sequence.
The first step is to calibrate the actuators of the input mirrors of each arm's cavity (NI and WI)
and of the beamsplitter mirror (BS) in a free swinging short Michelson configuration
using the main Virgo laser wavelength (1064\,nm) as length etalon.
In this configuration, from the interference fringes passing through the output photodiode,
the differential arm length $\Delta L(t)$ is measured using a non-linear reconstruction that has been described in~\cite{2011CQGra..28b5005A}.
Applying known excitations to the mirror actuators and looking at their effect on the reconstructed $\Delta L$, 
we can estimate the NI, WI and BS mirror actuator responses in meter per volt.
\ \\
Typical amplitude spectral densities of the reconstructed $\Delta L$ signal are shown in figure~\ref{fig:freemich_sens}.
The blue line represents the sensitivity of the measurements:
above 50~Hz, the sensitivity is dominated by the photodiode readout noise.
The green line shows an example when some sinusoidal excitations were applied to the different mirror actuators:
the applied excitations have signal-to-noise ratio of the order of few hundred up to 400~Hz. 
The NI, WI and BS mirror actuator responses have been measured up to 900~Hz.

\subsection{Calibration of the NE and WE mirror actuators}

Compared to initial Virgo, it is no more possible to calibrate directly the NE and WE end cavity mirror actuator 
with the free swinging Michelson technique, using asymmetric Michelson configurations,
because the sensitivity for this measurement is reduced due to the lower transmittivity of the input mirrors
and because the NE and WE actuators dynamics has been limited to only a Low Noise operation.
\ \\
The method therefore uses the NI and WI mirror actuator responses, $A_{in}$, as the reference to measure the NE and WE mirror actuator responses $A_{end}$.
With the full interferometer locked, one can compare the effect of known motions of the NI and WI mirrors on the dark fringe power
to the effect of the known excitations of the NE and WE mirrors. This comparison, that we call calibration transfer, allows to estimate the NE and WE mirror actuator responses.
\ \\
We can compute the transfer functions $TF_{in} = \mathcal{P}_{DC}/zCAL_{in}$ and\\
$TF_{end} = \mathcal{P}_{DC}/zCAL_{end}$ between the dark fringe signal $\mathcal{P}_{DC}$ and $zCAL_{in,end}$, 
the excitations injected on input or end mirrors of the arm cavities $zCAL_{in,end}$.
We can then extract the end mirror actuators response $A_{end}$ as:
\begin{eqnarray}
A_{end} & = &
                 A_{in} \times \frac{TF_{end}}{TF_{in}} \times \frac{R_{ITF,in}}{R_{ITF,end}}  \label{eqn:InToEnd}
\end{eqnarray}
where $R_{ITF,in}$ and $R_{ITF,end}$ are the interferometer optical response to the input and end mirrors motions
(the responses are almost the same, but with a 0.37\% difference in modulus and 10\,\mus\ difference in phase~\cite{calibnote_Repro2A}).
\ \\
The actuators response can be decomposed into two parts: the pendulum mechanical response and the electronic response.
NE and WE are suspended to a chain of anti-seismic suspension whose last stage can be modeled,
in the frequency range of interest, as a simple pendulum 
(a complex pole at $f_{pend} = 0.6\,\text{Hz}$ with a quality factor $Q = 1000$, following a simple $1/f^2$ behavior above 10~Hz).
Figure~\ref{fig:tfne} shows the NE mirror actuator response normalized by this simple mechanical response model.
The normalized modulus is mainly flat, within small deviations of $\pm3\%$ coming from the electronics/actuator response.
These small deviations are fitted (red curve) to provide the full actuator response model.
The superposed green areas indicate the $1\,\sigma$ statistical uncertainties as function of frequency estimated using the fit covariance matrix.
They are below 0.5\% in modulus and 5~mrad in phase up to 800~Hz.

\begin{figure}[tb]
  \begin{center}
    \includegraphics[angle=0,width=0.7\linewidth]{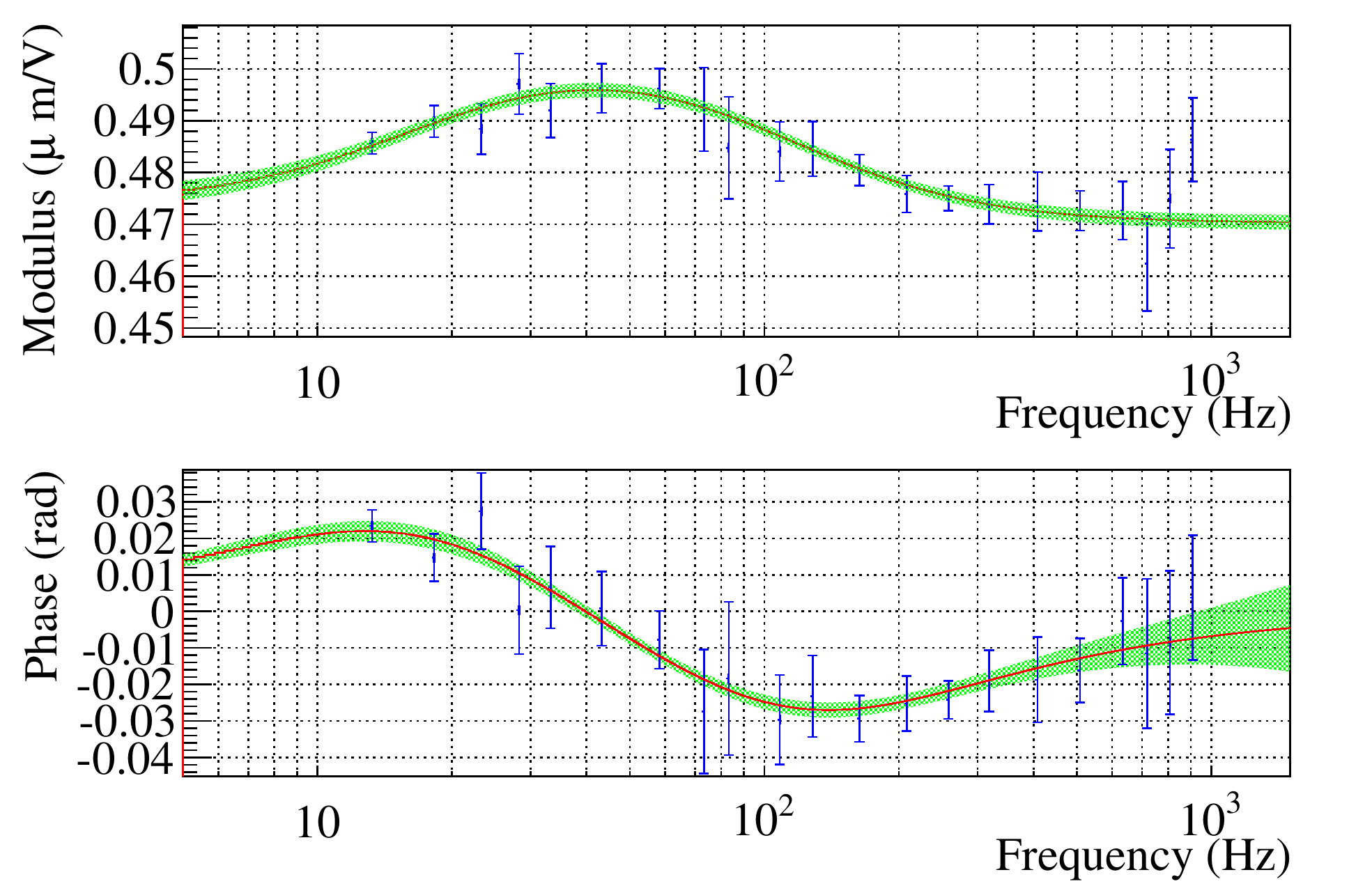} 
    \caption{NE mirror actuator transfer function normalized by the mechanical transfer function
      and with the measured delay subtracted for a better visualization.      
      Each blue point is an average over all the calibration measurements done at this frequency,
      with their $1\,\sigma$ statistical uncertainty.
      The red line is the fitted model for the (normalized) actuator response.
      The green filled area represents the $1\,\sigma$ statistical uncertainty as function of frequency.
    }
    \label{fig:tfne}
  \end{center}
\end{figure}

\subsection{Calibration of the marionnetta actuators}

The same transfer method is used in order to measure the Marionetta actuator response. 
We compare the effect, on the dark fringe power, of a known mirror excitation motion
to the effect of a known marionette actuator excitation.
The marionetta actuation response has been measured between 10~Hz and 200~Hz.
More details on this part of the calibration can be found in \cite{calibnote_Repro1A}.

\subsection{Calibration of the PR mirror actuator}

A similar transfer method is applied to measure the power-recycling mirror (PR) actuation response.
It uses the cavity made of the PR and WI mirrors, using BS as a folding mirror, 
to compare the effect, on the photodiode error signal used in the control loop, 
of the known WI motion to the effect of a known PR actuator excitation.
The PR actuation response has been measured between 10~Hz and 500~Hz.
More details on this part of the calibration can be found in \cite{calibnote_Repro1A}.

\begin{table}[tb]
\begin{center}
\small{
\begin{tabular}{|c|c|c|c|c|c|}
\cline{3-6}
\multicolumn{2}{ c|}{}                          & NE mirror       & WE mirror        &  BS mirror &  PR mirror \\
\hline
\multicolumn{2}{|c|}{Stat. uncertainty}         & 0.5\% (5 mrad)  & 0.5\% (5 mrad)   &  1\% (10 mrad) & 2\% (20 mrad)  \\
\hline 
\multirow{5}[0]{*}{\rot{Syst. uncert.}}   
 & BS,NI,WI calib                               & 0.4\% (0 mrad)  & 0.4\% (0 mrad)   & 0.2\% (3 mrad) & 0.4\% (0 mrad) \\
 & in to end transfer                           & 0 (0 mrad)      & 0 (0.5 mrad)     & --             & --             \\
 & WI to PR transfer                            & --              & --               & --             & 0.3\% (0 mrad) \\
 \cline{2-6}
 & $\Delta f$ in normalization                  & \multicolumn{4}{|c|}{0.2\% at 20 Hz and 0.04\% at 100 Hz  (0 mrad)} \\
 \cline{2-6}
 & Readout delay                                & \multicolumn{4}{|c|}{4~\mus}  \\
\hline
\multicolumn{2}{|c|}{Total uncertainty}         & 1.1\%            & 1.1\%            & 1.4\%         & 2.9\%       \\
\multicolumn{2}{|c|}{(linear sum)}              & 5 mrad           & 5.5 mrad         & 12 mrad       & 20 mrad     \\
\multicolumn{2}{|c|}{}                          & 4~\mus           & 4~\mus           & 4~\mus        & 4~\mus      \\
\hline                                                                                                            
\multicolumn{2}{|c|}{Validity range}            & 10-800 Hz        & 10-800 Hz        & 10-800 Hz     & 10-500 Hz   \\
\hline
\end{tabular}
\caption{Summary of the sources of statistical and systematic uncertainties on the mirror actuator models.
  For every source, the uncertainties on the modulus (phase) are given.
  The last lines give, for all the actuators, the sum of all the uncertainties reported in this table and their validity range.
  See text for details.
}
\label{tab:MirUncertainties}
}
\end{center}
\end{table}

\begin{table}[tb]
\begin{center}
\small{
\begin{tabular}{|c|c|c|c|c|}
\cline{3-5}
\multicolumn{2}{c|}{}                         & NE mario.         & WE mario.        & BS mario.      \\   
\hline                                                                                                 
\multicolumn{2}{|c|}{Stat. uncertainty}       & 2\% (20 mrad)     & 2\% (20 mrad)    & 0.5\% (5 mrad)  \\
\hline                                                                                                 
\multirow{5}[0]{*}{\rot{Syst. uncert.}}                                                               
 & BS,NI,WI calib                              & 0.4\% (0 mrad)   & 0.4\% (0 mrad)   & 0.2\% (3 mrad)  \\
 & in to end transfer                          & 0 (0 mrad)       & 0 (0.5 mrad)     &  --             \\
 & mir to mar transfer                         & 0.3\% (3 mrad)   & 0.2\% (4 mrad)   & 0 (1 mrad)      \\                                                                                               
\cline{2-5} 
 & $\Delta f$ in normalization                 & \multicolumn{3}{|c|}{0.2\% at 20 Hz and 0.04\% at 100 Hz (0 mrad)}  \\
\cline{2-5}
 & Readout delay                               & \multicolumn{3}{|c|}{4~\mus}                                        \\          
\hline                                                                                                 
\multicolumn{2}{|c|}{Total uncertainty}        & 2.9\%         &   2.8\%             & 0.9\%     \\ 
\multicolumn{2}{|c|}{(linear sum)}             & 23 mrad       &   24.5 mrad         & 9 mrad    \\       
\multicolumn{2}{|c|}{}                         & 4~\mus        &   4~\mus            & 4~\mus    \\       
\hline                                      
\multicolumn{2}{|c|}{Validity range}           & 10-100 Hz     &   10-100 Hz         & 10-80 Hz  \\                  
\hline                                                                               
\end{tabular}
\caption{Summary of the sources of statistical and systematic uncertainties on the marionette actuator models.
  For every source, the uncertainties on the modulus (phase) are given.
  The last lines give, for all the actuators, the sum of all the uncertainties reported in this table and their validity range.
  See text for details.
}
\label{tab:MarUncertainties}
}
\end{center}
\end{table}

\subsection{Uncertainties estimation}
\label{syst_uncertainty}

Each step of the calibration procedure contributes to the uncertainties in amplitude and phase of the actuator responses. 
Tables~\ref{tab:MirUncertainties} and~\ref{tab:MarUncertainties} give the total uncertainties
for each actuators response (mirrors and marionettes respectively) and the breakdown of the various contributions.

The first line gives the statistical uncertainties estimated after all the measurements have been combined together to
get the actuator response data and fit.
The systematic uncertainties are given in the following lines. 
For each step, data taken at different times have been averaged together.
However, for some of them, small time variations have been found: they are reported in the three next lines
for the corresponding steps (either the initial measurement of the BS, NI or WI actuation, 
either the calibration transfers to NE, WE, PR mirror or marionettes).
We found that, when normalizing the measured data by the simple pendulum model,
some bias was introduced by using the frequency of the Fourier transform bin center, which was not exactly the injected frequency.
This bias on the modulus, that decreases when the frequency increases, is also reported.
Finally, the uncertainty estimated on the photodiode readout timing is given.
The last lines of the tables summarize the total uncertainties obtained on the different actuator models and their validity range.
\ \\
For NE and WE mirrors, a main source of uncertainty is the statistical error, mainly coming from the calibration transfers.
Another main source of uncertainty on the modulus comes from the systematics due to the variation in time
of the actuators response measured in the free Michelson configuration.

\section{Reconstruction of the gravitational wave signal $h(t)$}
\label{recon}

Once the sensing chain and the control loop actuators are calibrated,
we can reconstruct the gravitational wave signal $h(t)$.
In this section, we report the principle of the $h(t)$ reconstruction,
the main method used to estimate the systematic uncertainties
and some consistency checks we performed.
Finally, the systematic uncertainties obtained for the online $h(t)$ version and for the two post-O2 reprocessings are summarized.
More details can be found in the Virgo notes~\cite{hrecnote_Repro1A, hrecnote_Repro2A}.

\subsection{Principle}
To compute $h(t)$, we remove from the dark fringe signal the contributions of the control signals by subtracting
the corrections applied to the mirrors and we correct for the interferometer optical transfer function.
This procedure does not correct the effect of control loops by applying a transfer function,
but by subtracting their contributions. Therefore, some of the residual control signals,
like the calibration lines (periodic excitations applied to the mirrors) are removed or reduced using this method.

Following the notations of figure~\ref{fig:itfconfig}, 
inputs are the measured dark fringe photodiode channel ($\mathcal{P}_{DC}$) 
and the correction channels  ($zC_i$) sent to the mirror and marionette actuators,
as well as the calibrated transfer functions for the photodiode readout and the different actuators.
As in initial Virgo, the computation is done in the frequency-domain using fast Fourier transforms of $20~$s with $10~$s overlap.
Calibration lines, applied as $zCAL_i$, are used to monitor the optical gain of the interferometer, 
that varies with the interferometer alignment for instance.

In addition, a frequency noise subtraction has been developed for the $h(t)$ reprocessings.
Indeed, it arose that frequency noise was still present in the dark fringe signal during O2 and could be subtracted
since it was also present in an auxiliary monitoring channel.
First, the transfer function from the frequency noise to $h(t)$ is computed and assumed constant over some time period (defined in \ref{hdiff}).
Then, the frequency noise contribution, estimated from the auxiliary monitoring channel and the transfer function,
is subtracted from $h(t)$.

The processing being done in the frequency domain, inverse fast Fourier transforms are used to obtain, in the time domain,
the final $h(t)$ channel provided at both 20000~Hz and 16384~Hz sampling frequencies.

\subsection{Main differences between the $h(t)$ versions}
\label{hdiff}
The Online $h(t)$ processing used the pre-O2 calibration models for the photodiode readout and the actuator responses.
The optical response of the interferometer was approximated by a simple pole characterized by the Fabry-Perot cavity finesse. 
The Online version used a fixed value of the finesse set to 455, the value measured in~2016 with an error of $\pm5\%$.\\
For the V1O2Repro1A $h(t)$ reprocessing, the three main changes have been 
to use the post-O2 improved actuation calibration models based on the whole calibration dataset, 
to correct for a timing bias of 116~\mus~\footnote{
This bias was due to a sign error for a 58~\mus\ correction in the reconstruction configuration.}
and to subtract the frequency noise. 
For this reprocessing, the transfer function from frequency noise to $h(t)$ has been fit to the measurements
and kept fixed for all the O2 data.
In addition, some improvements have been done in the reconstruction code to reduce some glitches due to communication issues in the control signals
that were found to happen during O2.\\
Finally, for the V1O2Repro2A $h(t)$ reprocessing, small adjustements of the calibration models were made 
(fix a bias of 0.37\% in the modulus of the end mirrors, and add $10~\mathrm{\upmu s}$ in the interferometer optical response to the end mirror motions).
The main modifications made on $h(t)$ processing have been:
\begin{itemize}
\item adapt the frequency noise subtraction transfer function every 500~s, monitoring the coupling of the frequency noise to $h(t)$,
\item adjust the cavity finesse every 10~s to the value extracted from the phase of some calibration lines measured 10~s earlier.
\end{itemize}
As an additional improvement, we have been less sensitive to glitches by using median values instead of 
average values for the optical gain and finesse.
Most of the results shown hereafter are computed from this V1O2Repro2A version.\\ 

Figure~\ref{fig:spectrumhoft} compares the amplitude spectral densities of the three versions of $h(t)$.
They are all very similar, but with some slight improvements of the reprocessing versions.
It is mainly visible between~15~Hz and~30~Hz and between 150~Hz and 200~Hz, 
where the sensitivity has been improved with respect to the online $h(t)$ reconstruction,
mainly thanks to the frequency noise subtraction.
At frequencies above 3~kHz, the Advanced Virgo data are contaminated by a significant amount of spectral and transient noise.
In the first reprocessing, the frequency noise subtraction was not well tuned at high frequency and added 
a slight excess of noise above 2~kHz that was no more present in the second reprocessing.

\begin{figure}[h!]
  \begin{center}
    \includegraphics[angle=0,width=1.1\linewidth]{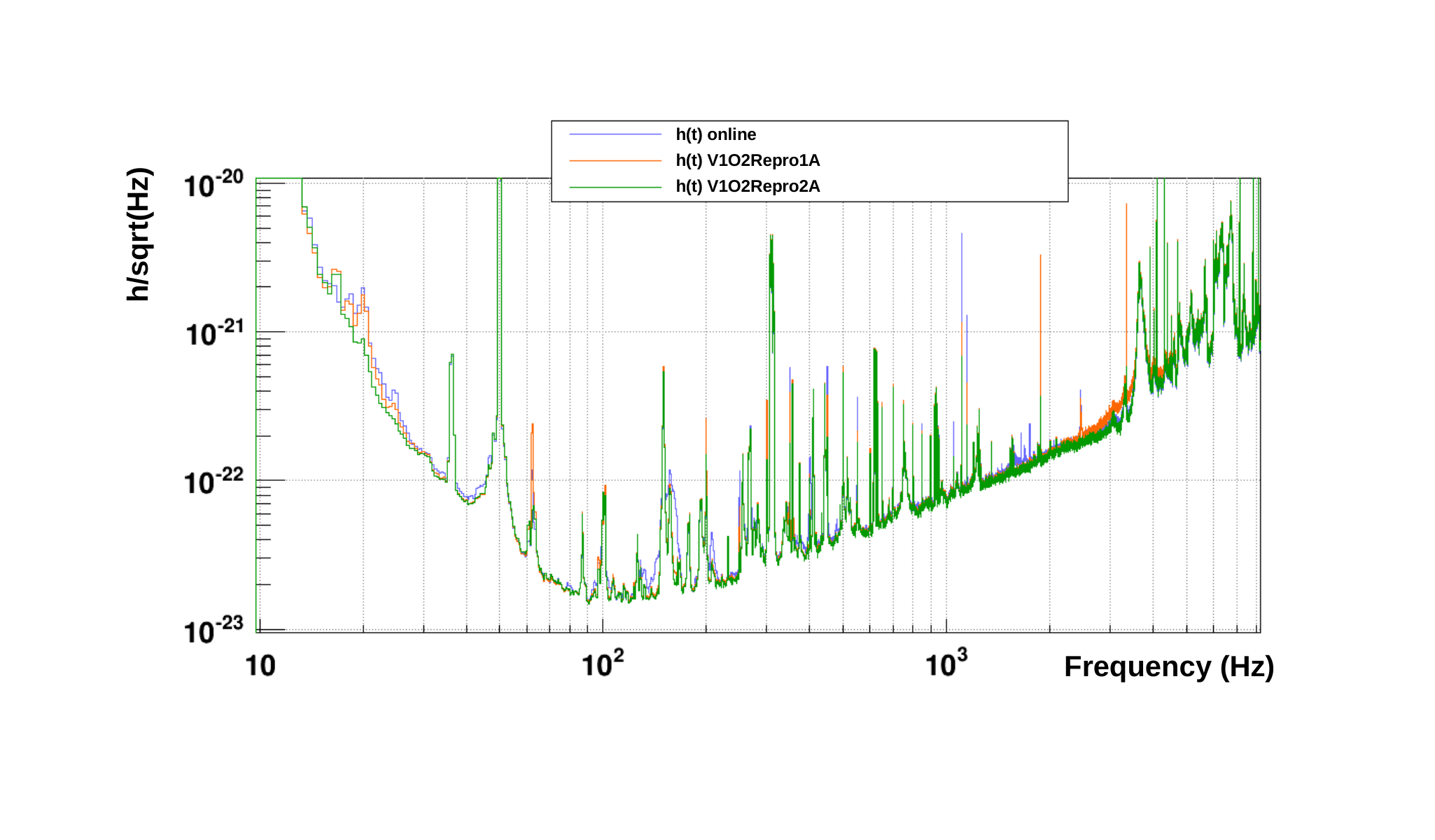} 
    \caption{Amplitude spectral density of $h(t)$ Online (blue), V1O2Repro1A (orange) and V1O2Repro2A (green).
    }
    \label{fig:spectrumhoft}
  \end{center}
\end{figure}

\subsection{Uncertainty budget}
To estimate the uncertainty on $h(t)$, including the possible bias of the reconstruction procedure, 
we have compared the reconstructed signal $h_{rec}$
to a calibrated $h_{inj}$ signal injected into the detector with the mirror electromagnetic actuators.
Figure~\ref{fig:waves2A} shows the amplitude and the phase of the transfer function $h_{rec}/h_{inj}$ where the
coherence between both signals was higher than 0.95.
Below~700~Hz, the comparison of $h_{rec}$ and $h_{inj}$ is within $\pm 4\%$ in amplitude and $\pm 35~\text{mrad}$ in phase,
as shown by the red lines in the figure.
In addition, the systematic uncertainty on the actuator model used to determine the reference $h_{inj}$ signal 
is $1.1\%$ in amplitude and 5~mrad in phase, and the systematic uncertainty on the timing is $20~\mathrm{\upmu s}$.

\begin{figure}[h!]
  \begin{center}
    \includegraphics[angle=0,width=0.8\linewidth]{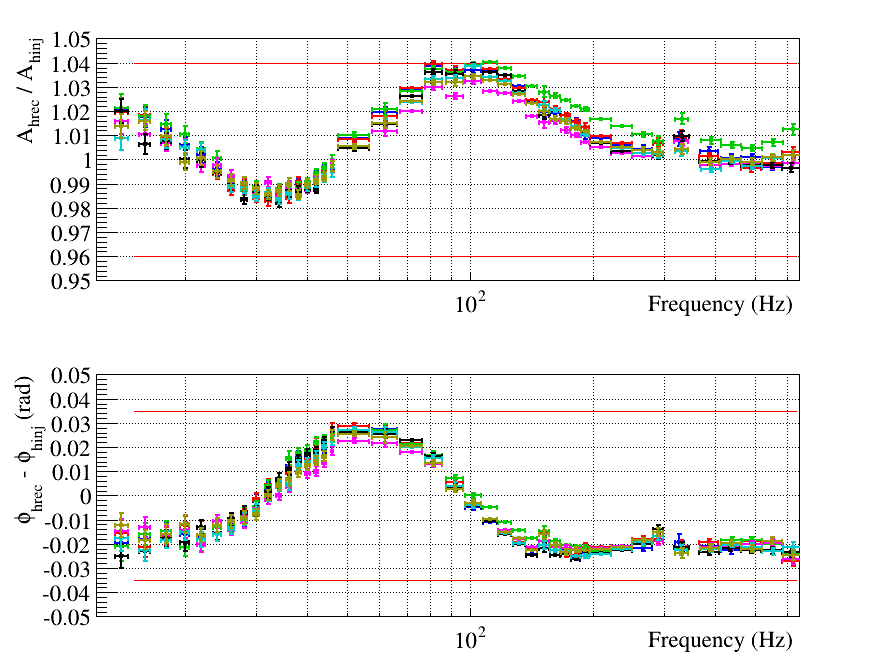}x
    \caption{Transfer function between the reconstructed $h(t)$ signal ($h_{rec}$) and the injected signals on NE mirror ($h_{inj}$).
             The different colors correspond to various sets of injections during the O2 run.
             The comparison of $h_{rec}$ and $h_{inj}$ is within $\pm4\%$ in amplitude and $\pm35$~mrad 
             in phase as shown by the red lines.}
    \label{fig:waves2A}
  \end{center}
\end{figure}

Due to the limited dynamic of the actuators, this method only applies up to $700~$Hz.
However the interferometer is basically kept free at high frequency since the contribution of the control signals
amounts to only $4\%$ of the dark fringe signal at $700~$Hz and goes down at higher frequencies.
Therefore, at high frequency, the $h_{rec}$ signal is just the dark fringe signal corrected 
by the cavity response and the photodiode electronic chain response. 
As an additional check, the shot noise level has been measured at high frequency 
as the difference between the power measured by the two dark fringe photodiodes.
As expected, it is flat, confirming that the bias introduced by the readout chain, and therefore the $h(t)$ uncertainty, is less than 4\%.
The $h(t)$ reconstruction is thus valid from 10~Hz up to the Nyquist frequency~8192~Hz
(for the~$h(t)$ channel sampled at~16384~kHz).
The systematic uncertainty on the $h(t)$ amplitude and phase for the online first and second reprocessings are reported in table~\ref{tab:hoft_uncertainties}.
~\\
Below 700~Hz, the frequency-dependent deviation seen in the figure is larger than the uncertainty on $h_{inj}$ coming from the actuators.
It must thus be a real bias in the reconstructed $h(t)$ channel. These deviations being in the frequency range where the photodiode signal
and the marionette and mirror control signals are all combined together to get $h(t)$, it is possible that small relative errors
in gain or phase of the different sensing and actuation models sum-up and give rise to such a structure.
Further investigations are on-going to really understand its origin.

\begin{table}[tb]
\begin{center}
\small{
\begin{tabular}{|c|c|c|c|}
\hline
  $h(t)$ version      & Amplitude uncertainty (\%)          & Phase uncertainty (rad)                    & Timing bias \\
  \hline
  Online              & $+14/-8$                & $100\times 10^{-3} + 2\pi f (20\times10^{-6})$  & $116\ \mathrm{\upmu s}$\\
  \hline
  V1O2Repro1A         & $\pm8$                  & $50 \times 10^{-3} + 2\pi f (20\times10^{-6})$  & 0\\
  \hline
  V1O2Repro2A         & $\pm5.1$                & $40 \times 10^{-3} + 2\pi f (20\times10^{-6})$  & 0\\
  \hline
\end{tabular}
\caption{Summary of the uncertainties estimated for the three versions of $h(t)$ reconstructed for the~O2 run.
For the Online version, the timing bias indicate that the reconstructed $h(t)$ was too late with respect to the {\it true} $h(t)$. 
It can either be corrected for when using the $h(t)$ channel or added linearly to the timing error.
The validity range is from 10~Hz to 8~kHz for the three versions.
}
\label{tab:hoft_uncertainties}
}
\end{center}
\end{table}

\section{Consistency checks with the photon calibrator (PCal)}

There are limitations in the way we can check the reconstructed $h(t)$ with respect to a differential arm length signal injected through
the electromagnetic actuators of the mirrors because the models used in the reconstruction of $h(t)$ are derived from the calibration 
procedure and an error on the actuator gains, larger than the computed uncertainties, would not be detected with this method.
\ \\
Performing a similar check but with another actuator, independently calibrated, allows to cross-check the results described in the previous sections.
This is provided by inducing mirror displacements with a photon calibrator (PCal).

\subsection{Principle of the photon calibrator}

The PCal is based on a $3~\mathrm{W}$ laser with a wavelength $\lambda = 1047~\text{nm}$ used to apply a force on
a mirror by radiation pressure. So far, in Advanced Virgo, one PCal was installed 
to push the WE mirror from the inner side of the Fabry-Perot cavity.
The force $F$ applied on the mirror is longitudinal:
\begin{equation} \label{eq:3}
F = \frac{2\cos (i)}{\text{c}}P_{ref}
\end{equation}
where $\text{c}$ is the speed of light in vacuum, $i$ the  angle of incidence of the laser on the
mirror and $P_{ref}$ the power of the PCal beam reflected by the WE mirror.
An excitation $\Delta F$ is applied by modulating the power of the PCal laser up to a few~kHz,
resulting in a mirror motion $\Delta x$ which verifies:
\begin{equation} \label{eq:4}
\Delta x = -\frac{1}{m}\frac{1}{(2\pi f)^2}\times \Delta F = -\frac{2\cos(i)}{\text{mc}}\frac{\Delta P_{ref}(t)}{(2\pi f)^2} 
\end{equation}
with $m$ is the mass of the mirror and $f$ the frequency of the sinusoidal force of amplitude $\Delta F$.
This equation holds for frequencies above 10~Hz, well above the resonant frequency of the pendulum.
This method enables us to compare an equivalent strain $h_{pcal}$ injected in the interferometer through
the PCal with the reconstructed signal $h_{rec}$. 
More details on the setup can be found in~\cite{bib:2015_Vnote_0013}.

\subsection{Sign of $h(t)$}

PCal injections can be used to establish the sign of $h(t)$, defined in equation~\ref{eq:1}, 
since we know that the variation $\delta L_{W}$ (i.e. $\delta L_y$) 
when the laser beam of the PCal pushes the mirror is the opposite of PCal power measurement $P_{ref}$.
Therefore, if the phase of the transfer function from the PCal power applied on WE mirror, $P_{ref}$, to $h(t)$ is equal to~0, 
equation~\ref{eq:1} is verified and thus the sign of $h(t)$ is correct. 
This assumes that the photodiode output is recorded as a positive quantity which can be easily checked
since there is a DC offset on the PCal laser power.
This is indeed what we measure.


\subsection{Partial confirmation of $h(t)$ systematic uncertainties with the PCal}

The transfer function of the reconstructed line amplitudes $h_{rec}$ over the injected line amplitudes with the PCal $h_{pcal}$ 
is shown on figure~\ref{fig:pcal}.
Since the calibration of the PCal beam power measurement was not stable during O2, with variations\footnote{
Post-run analysis have shown that the system was suffering from PCal beam polarization changes and some photodiode saturations.}
 of $\sim 20\%$,
the different transfer functions $h_{rec}/h_{PCal}$ measured during~O2 have been normalized in order to have a modulus of~1 
at the 356~Hz PCal calibration line.
In addition, since the PCal beam hits the center of the WE mirror, it excites the drum mode of the mirror. 
This modifies the simple mechanical response described in equation~\ref{eq:4}, 
adding a notch around 2~kHz (see~\cite{Accadia:2014fbz} for more details).
The left plot of figure~\ref{fig:pcal} shows the normalized amplitude ratio and the 2~kHz notch that was fitted on the data.
The right plot shows the phase difference, where the drum mode excitation has no impact 
as long as the measurements are below the notch frequency.
\ \\
From the normalized modulus plot, we cannot have an absolute verification of the amplitude of the reconstructed $h(t)$. 
However, the overall relative amplitude can be checked.
Once normalized, they lies within the $\pm 4\%$ error band that is used for the calibration uncertainty.
The important point is that the frequency-dependent deviation is similar to the one found with the electro-magnetic actuators shown in figure~\ref{fig:waves2A}.
This highlights that the deviation is a real bias in the reconstructed $h(t)$ signal.
This will be further investigated with future data taking.
\ \\
\begin{figure}[h!]
  \begin{center}
    \includegraphics[angle=0,width=1.1\linewidth]{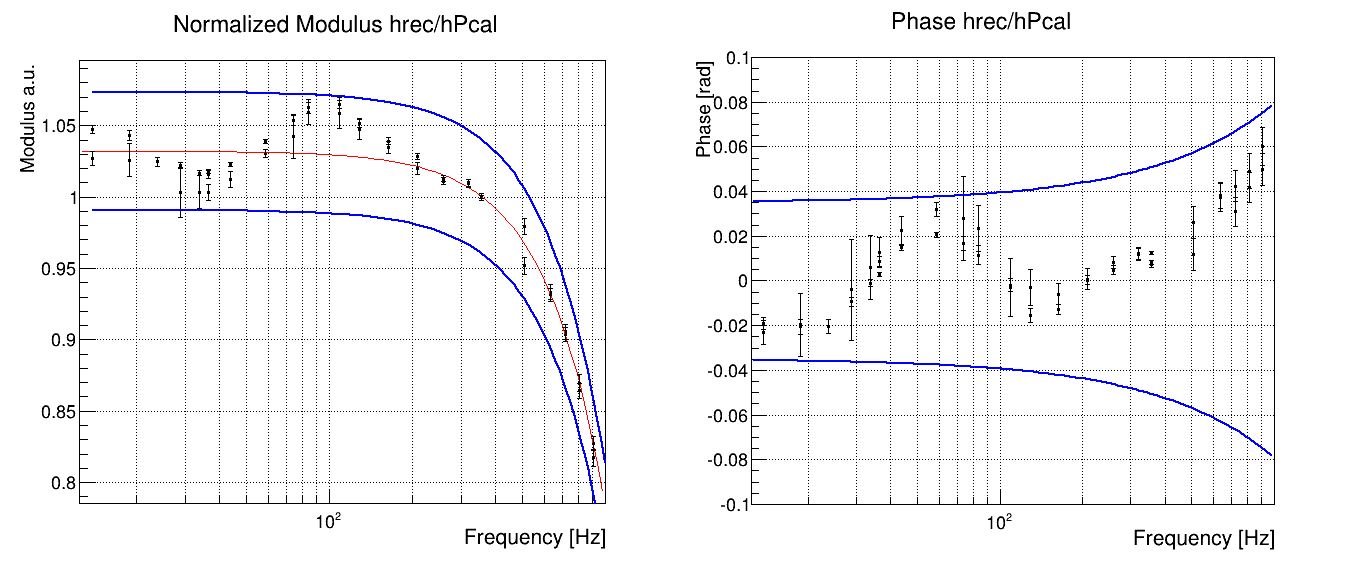} 
    \caption{Left plot represents the normalized transfer function between the reconstructed lines amplitude and the PCal injected lines amplitude. 
	The data points are fitted with a notch around 2~kHz and remain in the $\pm$4\% error band. 
	Right plot shows the phase of this transfer function with an error band of $\pm$35$\pm$2$\pi$f$\times$7e-3~mrad. 
}
    \label{fig:pcal}
  \end{center}
\end{figure}

The blue lines on the phase difference correspond to an error band of $\pm 35$ mrad uncertainty and a delay of $\pm 7~\mathrm{\upmu s}$ 
which is the sum of the timing uncertainties of the PCal photodiode ($3\,\mathrm{\upmu s}$) and of the dark fringe photodiode ($4\,\mathrm{\upmu s}$).
The timing of the PCal photodiode readout has been calibrated with a GPS clock signal, 
using the same method as described in section~\ref{lab:TimingAndSensing}.
The phase of the transfer function $h_{rec}/h_{PCal}$ lies within this error band.

To conclude, the comparison of the reconstructed $h(t)$ with the PCal injections has validated the sign of $h(t)$ 
and the estimated uncertainties on the phase of $h(t)$.
About the amplitude, while it cannot be used to validate the uncertainties, the fact that the frequency variation 
of the comparison are similar with both PCal and the standard mirror actuators gives extra confidence in the measurements.


\section{Conclusion}
We have described the Advanced Virgo calibration and the $h(t)$ reconstruction procedures used in the~O2 observation run in August~2017.
Using the interferometer laser wavelength as primary etalon and a calibration transfer procedure,
we could calibrate the BS, NE, WE and PR mirrors and marionetta actuators. They are important inputs
for the $h(t)$ reconstruction that subtracts from the dark fringe signal the control signals applied on the actuators.
The sources of systematic uncertainties have been described as well as the method to estimate the uncertainties on
the reconstructed $h(t)$ channel summarized in table~\ref{tab:hoft_uncertainties}, reaching 5.1\% in amplitude 
and 40~mrad and 20~\mus\ in phase and timing for the latest reprocessing version.

Following the first Virgo detections of gravitational wave signals during~O2, 
the LIGO and Virgo detectors are being improved and commissioned 
in preparation of the run~O3. With this three-detector network of improving sensitivity, there are strong prospects of many more
detections with increasing signal-to-noise ratio for the strongest events. 
In this context, the reconstruction uncertainties must be reduced down to the percent level in the coming years
in order not to contribute to the source parameter estimation uncertainties.

To achieve this goal, work is on-going to better calibrate the mirror actuators improving the photon calibrator setup
and installing a new calibration hardware, called Newtonian calibrator, that has shown good and consistent results in recent tests~\cite{bib:NCal}.
Due to the very tight schedule for the O2 run, time dedicated to calibration was very limited.
In order to reduce uncertainties and to refine the models used for the online reconstruction,
it is planned to keep the interferometer in a stable configuration with more calibration measurements before the start of future observation runs.

In the $h(t)$ reconstruction algorithm, studies to improve and adapt the optical models following the interferometer configuration are being pursued.
For O2, the bias found on the $h(t)$ signal has been included in the systematic uncertainties.
We plan to reduce it in the future by using a more accurate modelling of all actuator mirrors and optical responses,
and if needed by correcting the measured bias in the reconstructed $h(t)$ channel.


\section*{Acknowledgements}


The authors gratefully acknowledge the Italian Istituto Nazionale di Fisica Nucleare (INFN), the French Centre National de la Recherche Scientifique (CNRS) 
and the Foundation for Fundamental Research on Matter supported by the Netherlands Organisation for Scientific Research, 
for the construction and operation of the Virgo detector and the creation and support of the EGO consortium.

\section*{References}

\bibliographystyle{iopart-num}
\bibliography{references}

\end{document}